\documentclass[%
 reprint,
 amsmath,amssymb,
 aps,
]{revtex4-2}

\usepackage{graphicx}
\usepackage{dcolumn}
\usepackage{bm}
\usepackage{color}
\usepackage{makecell}

\begin{document}

\preprint{APS/123-QED}

\title{Active control of higher-order topological corner states \\in a piezoelectric elastic plate}
\author{Ze Ma}
\author{Yang Liu}
\author{Yu-Xin Xie}
\email{xyx@tju.edu.cn}
\author{Yue-Sheng Wang}%
\affiliation{%
 School of Mechanical Engineering, Tianjin University, Tianjin 300350, China}%



\date{\today}

\begin{abstract}
Different from the traditional bulk-edge correspondence principle, the discovery of higher-order topological states has generated widespread interest. In a second-order, two-dimensional elastic wave topological insulator, the fluctuation information can be confined to the corners, with the state being topologically protected. In order to better apply topological corner states, this paper designs a two-dimensional elastic plate with adjustable topological corner states by means of piezoelectric control capability. By selectively connecting negative capacitance circuits to piezoelectric sheets on the honeycomb elastic plate, the energy band can be flipped. The topological corner states at the 2$\pi$/3 corner were observed at the boundary of two different topological phase structures in the finite lattice with finite element software. The strong robustness of the topological corner states was verified by setting up defective control groups at the corner. In addition, the topological corner states of this piezoelectric elastic plate are discussed accordingly in terms of their tunability in frequency and position. The piezoelectric elastic plate is expected to provide a reference for the design of elastic wave local control and energy harvesting devices due to its adjustable topological corner states, which facilitate the application of topological corner states in practice. 
\begin{description}
\item[Keywords]
High-order topological insulators, Adjustable corner state, Piezoelectric elastic plate
\end{description}
\end{abstract}

\maketitle


\section{\label{sec:level1}introduction}

Topological Insulators (TIs) were originally studied in condensed matter physics\cite{von86q,hasan10tp,khanikaev13tp}. Their properties such as excellent suppression of fluctuation backscattering and strong robustness have attracted many researchers to explore them. Depending on the characterization of their topological invariants, they can be classified as quantum Hall insulators, quantum spin Hall insulators, and quantum valley Hall insulators\cite{khanikaev15t,he16s,foehr18sp,gu06v}. In recent years, topological devices have been extensively studied in the field of electromagnetic\cite{wang09g,wu15sc,khanikaev17t}, acoustic\cite{chen14a,xiao15g,lu16v,fleury16f,deng19s} and elastic\cite{wang15t,wu18t,chen18t,yang18t,chaunsali18s,huang20t,huang21r} waves. Their construction replies primarily on breaking the temporal and spatial inverse symmetry of the structure. Recently, higher-order TIs(HOTIs) have been theoretically predicted based on the extended bulk-edge correspondence principle\cite{benalcazar17q,schindler18h}. Unlike conventional first-order TIs, in second-order two-dimensional TIs, the (2-1)1D boundary state lacks topological protection, but maintains good robustness in the (2-2)0D corner state\cite{xue19a,peterson18q}. Experimental studies of HOTIs have also developed rapidly, as higher-order topological states have been better verified in topological metamaterials with quantized quadrupole\cite{serra18o,ni19o,xue19r}. By constructing a magnetic field to generate $\pi$ flux in the electromagnetic tetragonal lattice, the corner modes of the two-dimensional material can be observed accordingly\cite{peterson18q}. Based on an extension of the one-dimensional SSH model, the corner states of two-dimensional HOTIs are realized in lattices such as kagome, tetragonal and hexagonal lattices\cite{serra18o,xie18s,chen19c,zhang19s,kempkes19r,coutant20r,wu21c,ni19o,peterson18q,ezawa18h,xue19a,el19c,wang21h,xie20h,fan19e,chen21c}.

Active modulation of band gap frequencies and waves can be achieved by introducing some external factors into the phononic crystal or metamaterial\cite{wang20t}. In previous first-order elastic wave TIs, Gao et al. broke the mirror symmetry by rotating Y-shaped steel prisms on a substrate to achieve wide bandgap reconfigurable paths for topological valley transmission\cite{gao21r}. Zhang et al. designed a programmable lifting magnetic cavity to control the filling of the magnetic fluid to break the spatial inversion symmetry, and observed tunable topological valley transport in the experiment\cite{zhang19t}. By simulating the quantum valley Hall effect, Liu et al. achieved reconfigured topological waveguides via switching the sub-stable structure in the unit cell\cite{liu21t}. Taking advantage of the electrodynamic deformation properties of the dielectric elastomer, Chen et al. have investigated the active control of pseudospin boundary states in the soft-film type metamaterial, widening the frequency operating range\cite{chen21t}. By using the piezoelectric control technique, Li et al. controlled the appearance and disappearance of the double Dirac cone by adjusting negative capacitance circuits connected to piezoelectric sheets, and verified the strong robustness of the topological waveguide in the experiments\cite{li20a}. Piezoelectric shunt technology was first proposed by Forward, as well as being well developed for active control in the field of sound and vibration\cite{forward79e,casadei09b,airoldi11d,chen14b,sugino18m,liu21P,liu21M}. By connecting a negative capacitance circuit to the piezoelectric sheet of the metamaterial plate, it can be regarded as an element of adjustable stiffness. The local resonant frequency of the resonator can be controlled by adjusting the negative capacitance circuit\cite{airoldi11d,liu21M}.

Following the study of HOTIs for electromagnetic and acoustic waves, the corner states of elastic wave HOTIs have recently begun to be explored as well. Fan et al. designed elastic honeycomb phonon crystal plates with different intercellular and intracellular coupling strengths and experimentally observed the presence of topological corner states\cite{fan19e}. Based on the study of acoustic Wannier-type HOTIs, Wu et al. have experimentally verified two-dimensional second-order corner states in an elastic Kagome lattice plate\cite{wu20p}. Chen et al. observed mechanical topological corner states at the interface between two structures, topologically trivial and topologically non-trivial, by cyclically mounting steel bolts on the aluminum plate\cite{chen21c}. However, there lacks research on the tunable corner states of elastic-wave two-dimensional HOTIs. Thus, inspired by the structure of the one-dimensional local resonance tunable topological states by Liu et al.\cite{liu21P} we attached negative capacitance circuits to piezoelectric sheets on the inner or outer supports of the hexagonal lattice. In this way, the topological corner states of the elastic plate can be actively tuned. By adjusting the size of the capacitor parameter, the operating frequency range of its topological corner states can be widened. And by selectively connecting negative capacitance circuits to the piezoelectric sheets of the holder, active control of the topological corner states position can be achieved.

\section{Structural design and topological energy band inversion of the piezoelectric elastic plate}
\begin{figure*}[!htb]
	\centering
	\begin{minipage}[c]{.4\textwidth} 
		\centering\includegraphics[scale=0.45]{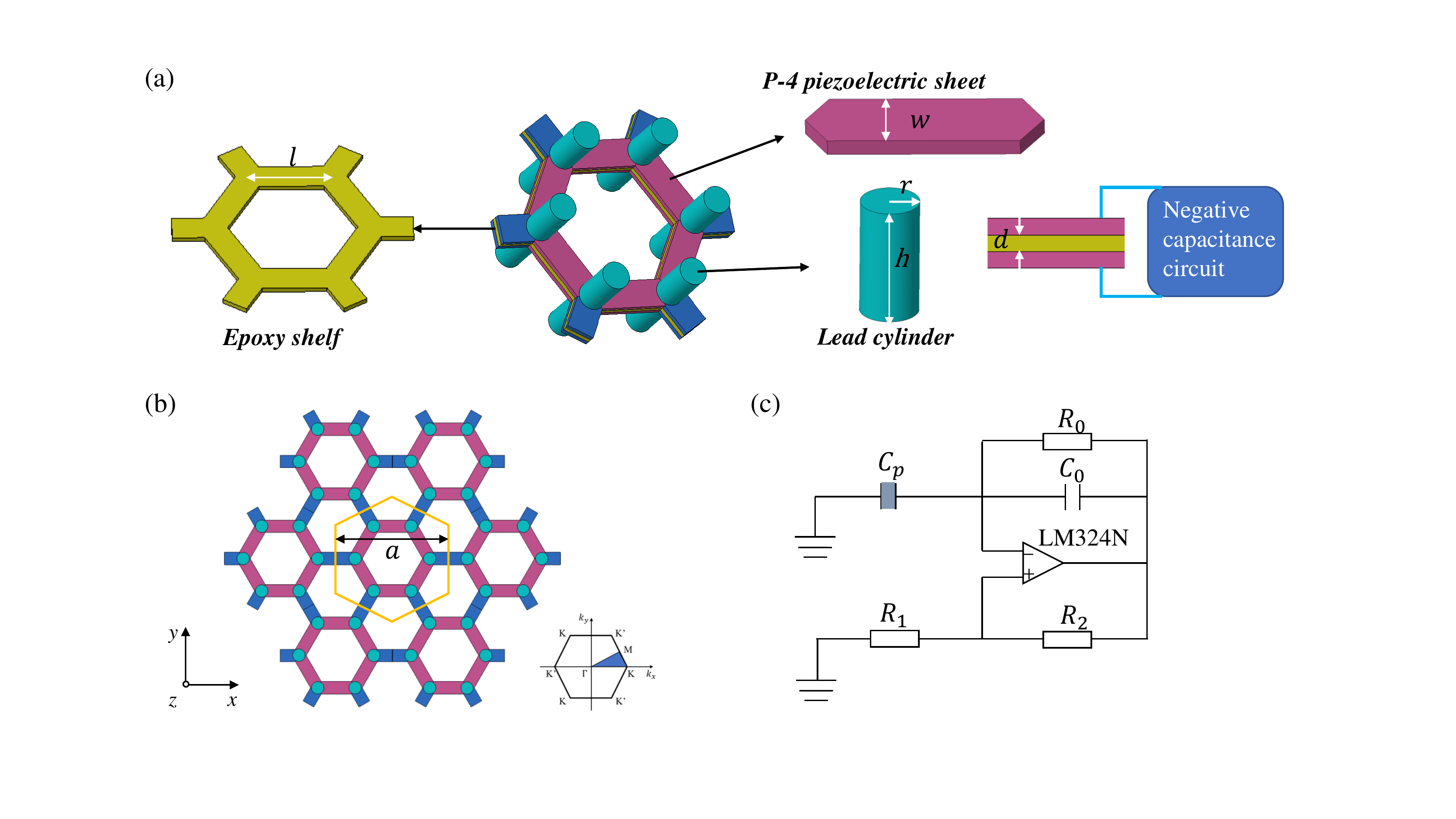}
	\end{minipage}
	\hfill 
	\begin{minipage}[c]{.3\textwidth} 
		\centering
		\caption{(a) Schematic of the unit cell structure of a piezoelectric elastic plate. The yellow honeycomb support serves as the flexible plate substrate. P-4 piezoelectric sheets are attached to the top and bottom of the holder, with the piezoelectric sheets angled at 120$^{\circ}$ at both ends. The parts marked in purple are the internal hexagonal piezoelectric sheets, and the external ones are marked in blue. Cylindrical magnet oscillators are glued to the top and bottom of the six corners of the honeycomb elastic plate. Negative capacitance circuitry is connected to the surface of the piezoelectric sheets. (b) Composite unit structure of piezoelectric elastic plates. The part enclosed by the yellow line frame is taken as the unit cell. (c) Schematic of the negative capacitance circuit.}
		\label{tu1}
	\end{minipage}
\end{figure*}

As shown in Fig. \ref{tu1}a, the piezoelectric elastic plate designed in this paper is composed of the substrate, the cylindrical oscillators and the piezoelectric sheets. The material of the substrate is epoxy resin in the shape of a continuous hexagonal lattice with thickness $d=2$ mm, side length $L=15$ mm and width $w=5$ mm. At the top and bottom of the substrate are pasted P-4 piezoelectric sheets of the same thickness as the substrate. The six corners of the hexagonal lattice have cylindrical magnet oscillators glued above and below them, where the oscillators have radius $r=2.5$ mm and height $h=5d$. The material parameters of the magnet, the epoxy and the P-4 piezoelectric sheet are shown in Tab. \ref{tab1}. The multi-cell structure of the piezoelectric elastic plate can be seen in Fig. \ref{tu1}b. The part of the diagram enclosed by the yellow wire frame is taken as the unit cell, with the length of the cell $a=45$ mm. In order to distinguish between the piezoelectric pieces, we marked the inner and outer hexagonal piezoelectric pieces as purple and blue respectively. The negative capacitance circuit in the active control system is illustrated in Fig. \ref{tu1}(c). It includes capacitor $C_{p}$, compensation resistor $R_{0}$, operational amplifier LM324N, fixed resistor $R_{1}$ and sliding resistor $R_{2}$. Tab. \ref{tab2} provides the parameters for each device of the negative capacitance circuit. The equivalent bending stiffness of the piezoelectric sheet can be controlled using a negative capacitance circuit connected to the P-4 piezoelectric sheet. A topological phase change of the elastic plate is thus achieved. 

\begin{table*}[!ht]
	\setlength{\abovecaptionskip}{0.1cm} 
	\centering
	\caption{\textbf{Material parameters for piezoelectric elastic plate.}}
	\label{tab1} 
	\resizebox{\linewidth}{!}{
		\begin{tabular*}{\hsize}{@{\extracolsep{\fill}}c c c c c c } 
			\hline
			Material & \makecell[c]{Modulus \\$E$(Pa)}  & \makecell[c]{Density \\$\rho$(kg/m$^{3}$)} &  \makecell[c]{Compliance \\coefficient \\$s^{E}_{11}$ (m$^{2}$/N)} & \makecell[c]{Piezoelectric 
				strain \\coefficient \\$d_{31}$(C/N)} & \makecell[c]{Permittivity \\$\varepsilon_{33}^{T}$(F/m)}\\
			\hline
			Magnet& 41$\times$10$^{9}$ & 7400 & ...&...&...\\
			\hline
			Epoxy& 3$\times$10$^{9}$ & 1180 & ...&...&...\\
			\hline
			\makecell[c]{P-4 \\piezoelectric sheet}& 8.83$\times$10$^{10}$ & 7450 & 1.2$\times$10$^{-11}$&-1$\times$10$^{-10}$ & 1.2$\times$10$^{-8}$\\
			\hline
	\end{tabular*}}
\end{table*} 
\begin{table}[!h]
	\setlength{\abovecaptionskip}{0.1cm} 
	\centering
	\caption{\textbf{The parameters of the negative capacitance circuit.}}
	\label{tab2} 
	\resizebox{\linewidth}{!}{
		\begin{tabular*}{\hsize}{@{\extracolsep{\fill}}c c c c c c } 
			\hline
			$C$(pF) &$C_{p}$(pF) &$R_{2}$(k$\Omega$) &  $R_{1}$(k$\Omega$) & $R_{0}$(k$\Omega$) & Operational 
			amplifier\\
			\hline
			9.684& 8.155& 51.54 & 68&2000&LM324N\\
			\hline
	\end{tabular*}}
\end{table}

In the case when the piezoelectric sheet is connected with a negative capacitance circuit, we derive the equivalent elastic module of the piezoelectric sheet as follows. The $x$, $y$ and $z$ directions are denoted as $1$, $2$ and $3$ respectively. Since the piezoelectric sheet is subjected to an electric field along the $z$-axis only, the intrinsic equation of the piezoelectric material in the plane stress state can be expressed as\cite{li20a,liu21P,liu21M} 
\begin{equation}
	\begin{aligned}
		S_{1}&=s_{11}^{E}T_{1}+d_{31}E_{3},\\
		D_{3}&=d_{31}T_{1}+\varepsilon_{33}^{T}E_{3},
	\end{aligned}
\end{equation}
where $S_{1}$ and $T_{1}$ are the strain and stress along the $x$-direction, $s_{11}^{E}$ represents the coefficient of flexibility under a constant electric field, $d_{31}$ and $\varepsilon_{33}^{T}$ are the piezoelectric and dielectric constants, respectively, and $E_{3}$ and $D_{3}$ describe the electric field strength and potential shift along the $z$-direction.

At a constant strain, the intrinsic capacitance of a piezoelectric sheet can be denoted as
\begin{equation}
	C_{p}=\varepsilon_{33}^{T}A_{s}h_{p}^{-1},
\end{equation}
where $A_{s}$ and $h_{p}$ indicate the area and thickness of the piezoelectric sheet respectively. Then, introducing the complex impedance $Z$, the relationship between strain and stress can be derived as
\begin{equation}
	S_{1}=\left(s_{11}^{E}-\frac{sZd_{31}^2A_{s}}{(1+sZC_{p})h_p}\right)T_{1},
\end{equation}
where $s$ is the Laplace parameter.

The equivalent modulus of a piezoelectric sheet connected to a circuit with negative capacitance can be derived from equation (3) as follows:
\begin{equation}
	E_{p}=\frac{h_{p}(1+sZC_{p})}{h_{p}s_{11}^{E}(1+sZC_{p})-sZd_{31}^{2}A_{s}}.
\end{equation}
Its complex impedance $Z$ is expressed as
\begin{equation}
	Z=-\left({\alpha s C_{p}}\right)^{-1},
\end{equation}
where $\alpha={R_{2}C}({R_{1}C_{p}})^{-1}$.

Therefore, we can make the elastic modulus of the piezoelectric sheet varied by adjusting the parameter $\alpha$, i.e., regulating the joint stiffness of the piezoelectric elastic sheet.

\begin{figure*}[!ht]
	\centering
	\begin{minipage}[c]{0.5\textwidth} 
		\centering\includegraphics[scale=0.7]{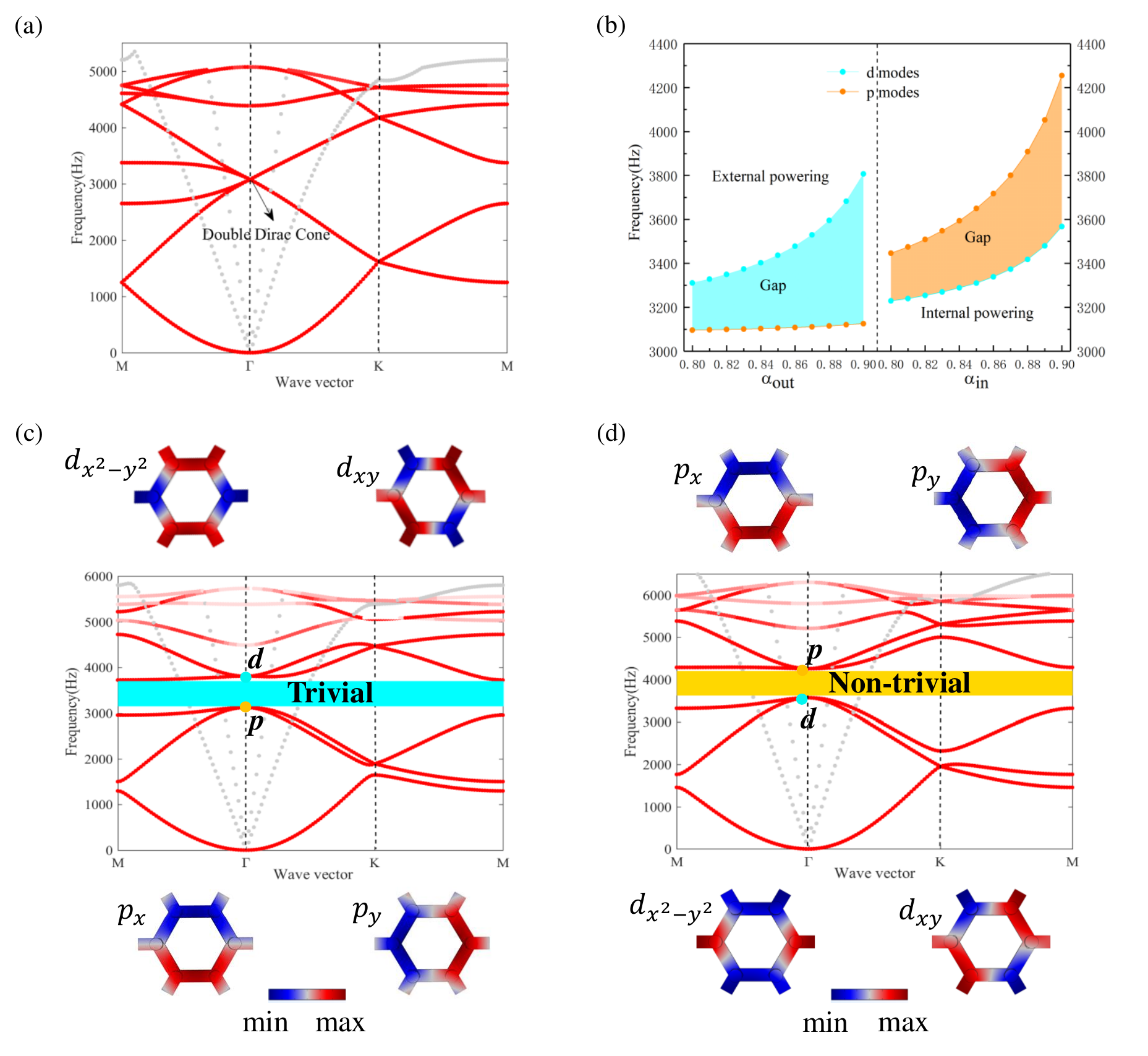}
	\end{minipage}
	\hfill 
	\begin{minipage}[c]{.22\textwidth} 
		\centering
		\caption{(a) The unit cell dispersion curve for the piezoelectric sheet without the negative capacitance circuit connected. (b) Curves for the variation of band gap width with parameter $\alpha$ when the honeycomb elastic plate is internally or externally charged. (c), (d) Band structures and their intrinsic mode patterns when the external and internal piezoelectric sheets are charged for $\alpha$= 0.9.}
		\label{tu2}
	\end{minipage}
\end{figure*}

Next, the band structures of the unit cell were analyzed using COMSOL Multiphysics software. The stiffness factor is the same for each holder when neither the inner nor outer piezoelectric sheet is connected to the negative capacitance circuit. In Fig. \ref{tu2}(a), the four bands degenerate at point $\Gamma$, forming a double Dirac cone. In the band structures we are only concerned with the out-of-plane modes (marked in red), and the in-plane modes (marked in grey) are not considered. The stiffness factor of the corresponding elastic support is enhanced when an internal or external piezoelectric sheet is connected to a negative capacitance circuit. As a result, the double Dirac cone disappears and a band gap forms at the corresponding location nearby. As can be seen in Fig. \ref{tu2}(b), the forbidden band width widens as the parameter $\alpha$ in the negative capacitance circuit increases. The $d$ modes in the band structure are above the band gap and the $p$ modes are below when the external piezoelectric sheets are connected to negative capacitance circuits. Instead, as the internal piezoelectric sheets are connected to negative capacitance circuits, the $d$ and $p$ modes are interchanged in their distribution above and below the band gap. A transition from a trivial state to a non-trivial state occurs in the topological phase during this period. Specifically, external powering produces the trivial bandgap (Fig. \ref{tu2}(c)), while internal powering corresponds to the non-trivial bandgap (Fig. \ref{tu2}(d)). The eigenmode diagrams are plotted for the high symmetry point $\Gamma$ in the upper and lower bands forming the band gap. They are shown accordingly at the top and bottom of the band structures diagram. In this case, the $d$ modes represent quantized four polarization. The mode distribution of $d_{x^{2}-y^{2}}$ is symmetric about the $x$ and $y$ axes, while the mode of $d_{xy}$ is antisymmetric. The $p$ modes represent quantized couple polarization. $p_{x}$ and $p_{y}$ modes are antisymmetrically distributed about the $y$ and $x$ axes respectively\cite{yang18t}.

\section{Discussion of topological corner states}
\begin{figure*}[!ht]
	\centering
	\renewcommand{\figurename}{Fig}
	\begin{minipage}[c]{0.15\linewidth}
		\centering
		\includegraphics[scale=0.42]{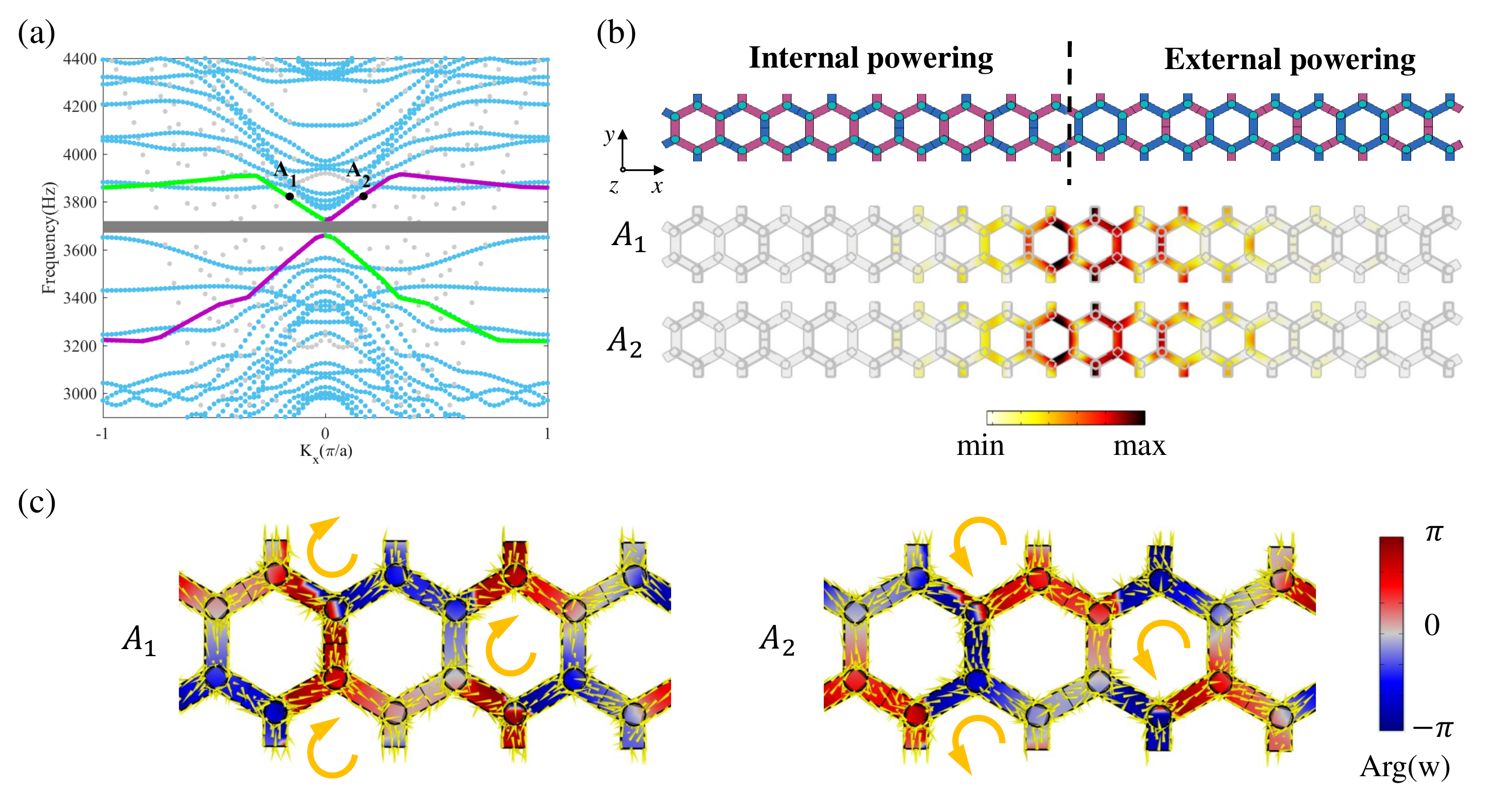}	
	\end{minipage}
	\hfill
	\begin{minipage}[c]{0.22\linewidth}
		\caption{(a) Dispersion relation diagram of the supercell. (b) Schematic of the structure of the supercell and the corresponding vibrational modes at points $A_{1}$ and $A_{2}$. The color bar represents the size of the absolute value of the displacement in the $z$-direction of the model. (c) Cloud field plot of the phase distribution and energy flow arrows for the $z$-direction eigenstates at the modal interface of points $A_{1}$ and $A_{2}$.}
		\label{tu3}
	\end{minipage}
\end{figure*}
Further, the supercell dispersion relation of the piezoelectric elastic plate was analyzed. This supercell consists of six internally powered unit cells on the left and six externally powered unit cells on the right. The left and right ends of the supercell are the free boundary condition, and the Floquet-period boundary condition along the $y$-axis is imposed at its upper and lower ends. The dispersion relation obtained from the computational analysis is shown in Fig. \ref{tu3}(a), with the grey and light blue marked points being the in-plane and out-of-plane modes respectively. We focus only on the out-of-plane modes, with two clear topological boundary states (green and purple) in the middle part, corresponding to the chiral propagation model generated by the pseudo-spin Hall effect. The topological boundary vibration mode diagrams corresponding to points $A_{1}$ and $A_{2}$ are shown in Fig. \ref{tu3}(b), where it can be observed that the displacement is greatest at the intermediate interface. To better describe the two topological pseudo-spin states, we show the energy flow arrows (yellow) at the interface as in Fig. \ref{tu3}(c). In the boundary mode at points $A_{1}$ and $A_{2}$, the energy flow appears in clockwise and counterclockwise directions respectively. Meanwhile, we plot the phase field of the eigenstates at the boundary as the cloud field. It can be observed that the phases of the two boundary modes at the same position are opposite to each other. This suggests that the two pseudospin boundary states have propagation modes in opposite directions. But such boundary states differ from the gapless edge states of fermion pseudospin protected by time-reversal symmetry\cite{kane05q}. In the middle of the two edge states of this model band structure, there is a grey forbidden band. The band gap is created by the breaking of $C_{6}$ crystal symmetry at the boundary of different topological phase structures\cite{chen21c}. We try to find the topological corner state by referring to the frequency distribution of this band gap.

\begin{figure}[!ht]
	\centering
	\renewcommand{\figurename}{Fig}
	\begin{minipage}{1\linewidth}
		\centering
		\includegraphics[scale=0.52]{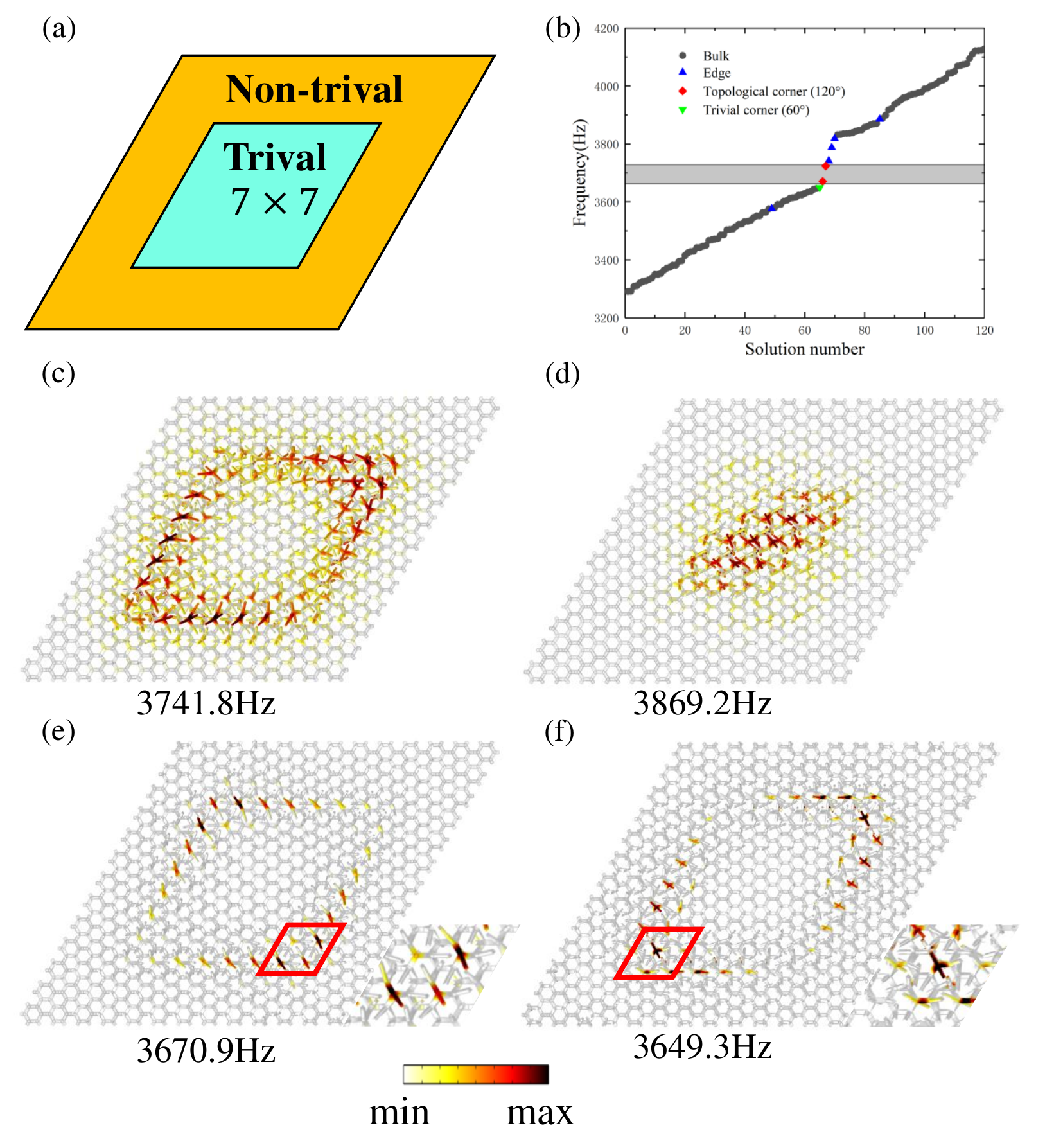}	
	\end{minipage}
	\caption{Calculated eigenmodes of the piezoelectric elastic plate.(a) Schematic diagram of a diamond-shaped finite lattice. The interior is the externally charged trivial phase structure (cyan) and the periphery is the internally charged non-trivial phase structure (yellow). (b) Numerical calculation of the eigenfrequencies obtained. The red rhombus represents the topological corner state, while the grey circle, blue square triangle and green inverted triangle represent the bulk, edge and trivial corner states respectively. (c) Edge state at 3741.8 Hz. (d) Bulk state at 3869.2 Hz. (e) Topological corner state at 3670.9 Hz. (f) Trivial corner state at 3649.3 Hz. The color bar indicates the absolute magnitude of the displacement in the $z$-direction.}
	\label{tu4}
\end{figure}

In order to analyze the topological corner states of the piezoelectric elastic plate, we constructed a 13$\times$13 diamond-shaped finite lattice model as shown in Fig. \ref{tu4}(a). The parameter $\alpha$ for the negative capacitance circuit connected to the piezoelectric sheet is set to 0.9. The external bracket of the 7$\times$7 unit cell inside the model is charged, corresponding to the trivial phase structure analyzed above. The outer unit cell is then a non-trivial phase structure. The two different phase structures thus form a diamond-shaped boundary. The eigenfrequencies of the model were calculated in COMSOL Multiphysics software and the resulting eigenfrequency spectrum is shown in Fig. \ref{tu4}(b). The presence of bulk states (grey circles), edge states (blue square triangles), topological corner states (red rhombuses) and trivial corner state (green inverted triangle) can be observed. The corresponding vibration mode diagrams for the edge and bulk states are plotted in Figs. \ref{tu4}(c, d). We are primarily concerned with topological corner states that occur in the bandgap range, which are topologically protected. The topological corner state mode diagram corresponding to the eigenfrequency of 3670.9 Hz is shown in Fig. \ref{tu4}(e). It can be observed in the diagram that the larger absolute values of out-of-plane displacement $\lvert w \rvert$ are concentrated at the 2$\pi$/3 corners of the bend at the rhombic boundary. The corner state corresponding to 3649.3 Hz below the band gap is the trivial corner state. In the corresponding mode vibration diagram, the larger displacements occur at the $\pi$/3 corners of the boundary. In analogy to the theoretical explanation in the higher-order topology of electromagnetic waves, there is a zero mode (topological corner mode) at each uncoupled waveguide\cite{noh18t}. Each zero mode holds one topological charge ($+$ or $-$). $N_{+}$ and $N_{-}$ are used to indicate the numbers of eigenstates of the chiral symmetric operator $\Pi$ with topological charges $+$ and $-$, respectively\cite{fan19e}. There are three zero modes at the 2$\pi$/3 corner, two of which contain the same topological charge and one of which contains the opposite topological charge. As a result, the topological index $\mathcal N=\lvert N_{+}-N_{-}\rvert=1\neq0$, representing the topological corner state. The details can be found in Appendix B. While the topological charge is positive at the two zero modes at $\pi$/3, that at the other two zero modes is negative, so the topological index $\mathcal N$=0. These theories can help to explain the phenomenon that the topological corner states of the present honeycomb piezoelectric elastic plate appear only at the 2$\pi$/3 corners.

\begin{figure*}[!ht]
	\centering
	\renewcommand{\figurename}{Fig}
	\begin{minipage}{0.6\linewidth}
		\centering
		\includegraphics[scale=0.6]{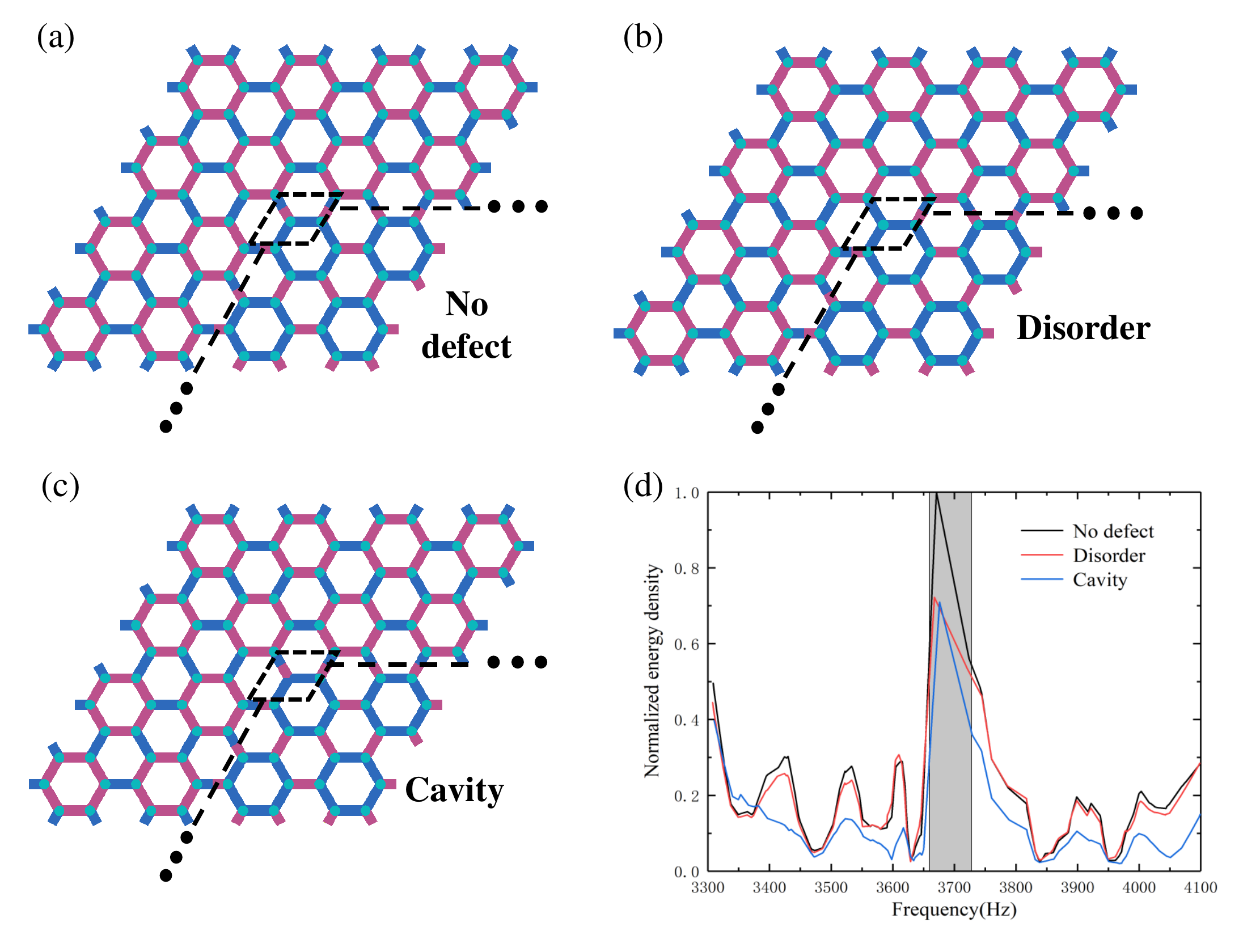}	
	\end{minipage}
	\hfill 
	\begin{minipage}[c]{.25\textwidth} 
		\centering
		\caption{(a) Partial diagram of the finite lattice at the 2$\pi$/3 corner. The purple holders indicate that a negative capacitance circuit is connected. Non-trivial phase structures outside the corner and trivial phase structures inside the corner. (b) Disturbed finite lattice model. A holder at the corner is not connected to the negative capacitance circuit, creating a disturbance to the topology model. (c) Cavity finite lattice model. The absence of two oscillators above and below the support at the corner forms a cavity defect topology model. (d) Normalized energy spectrum curves for the non-defective and defective models. The grey band represents the boundary state band gap of the supercell. The black wire represents the non-defective model, with the red and blue wires representing the disturbed and cavity defect models respectively.}
		\label{tu5}
	\end{minipage}
\end{figure*}

Next, we verified the strong robustness of the piezoelectric elastic plate topological corner states. The $2\pi/3$ corner of the interface between the different phase structures in the finite lattice is displayed magnified. Fig. \ref{tu5}(a) shows the defect-free state, while Fig. \ref{tu5}(b, c) represent the disturbed and cavity models, respectively. A point at the $2\pi/3$ corner of the boundary is chosen to plot the normalized frequency energy spectrum as shown in Fig. \ref{tu5}(d). There is a clear high-energy peak in the grey band gap of the defect-free structure, which corresponds to the presence of the topological corner states. In the presence of disturbance and vacancy defects, the frequency interval with the higher energy peak remains confined to the band gap. And the average elastic energy density peaks at the corner of the defective structures are close to the peak in the case of the defect-free structure, with less dissipation. The resulting analysis indicates that the topological corner state of the piezoelectric elastic plate is insensitive to small defects with strong robustness.

\begin{figure}[!ht]
	\centering
	\renewcommand{\figurename}{Fig}
	\begin{minipage}{1\linewidth}
		\centering
		\includegraphics[scale=0.5]{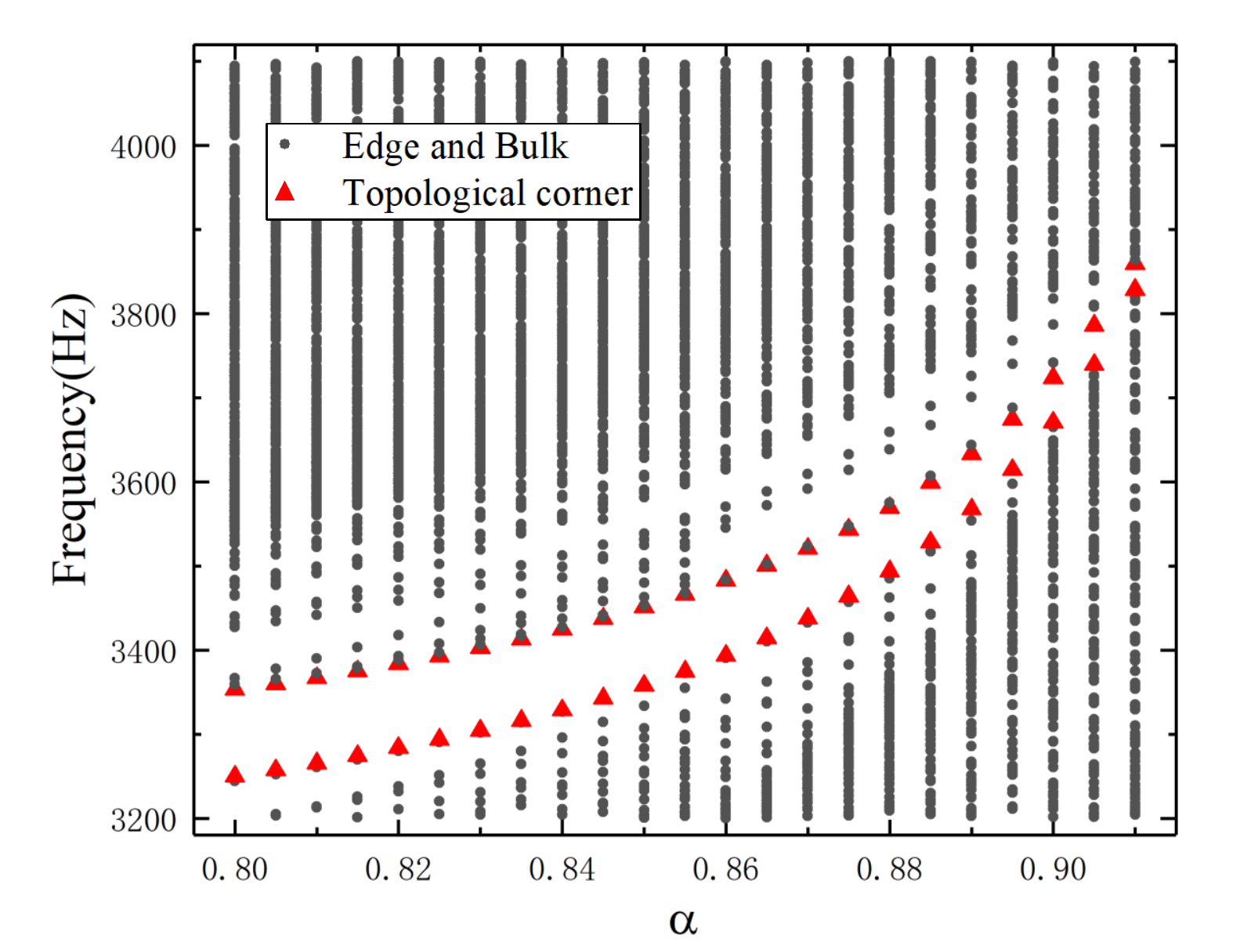}	
	\end{minipage}
	\caption{Variation of the finite lattice frequency spectrum with the piezoelectric parameter $\alpha$. The grey dots indicate the bulk and edge states and the red triangles indicate the topological corner states.}
	\label{tu6}
\end{figure}

\begin{figure}[!ht]
	\centering
	\renewcommand{\figurename}{Fig}
	\begin{minipage}{0.96\linewidth}
		\centering
		\includegraphics[scale=0.48]{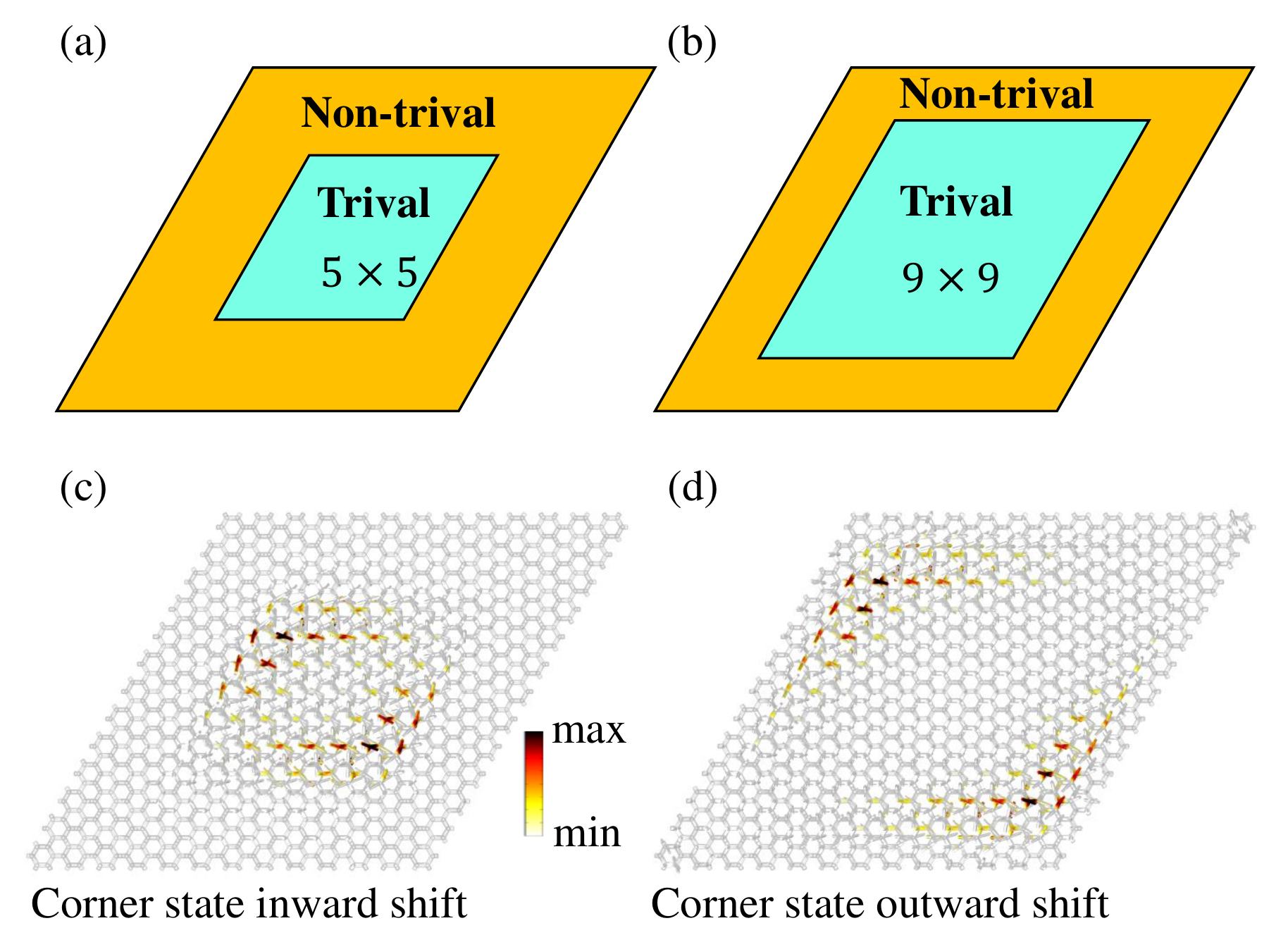}	
	\end{minipage}
	\caption{(a)-(b) Schematic of the corner state inward and outward shift model for the finite lattice (13*13 cells). (c)-(d) Inward and outward shifted vibration mode graphs for corner states.}
	\label{tu7}
\end{figure}

To emphasize the frequency tunability of the topological corner states of this piezoelectric elastic plate, we investigated the topological corner state spectrum variation of its finite lattice as shown in Fig. \ref{tu6}, with the piezoelectric parameter $\alpha$ as a variable. As $\alpha$ increases, the stiffness of the holders connected to negative capacitance circuits rises and the topological corner states of the piezoelectric elastic plate appear with an accordingly higher frequency. Also, as the parameter increases, the two topological corner states become closer and closer. It means that the piezoelectric elastic plate can achieve topological corner states in multiple frequency ranges, and the way to go is to tune the piezoelectric parameter. Besides, compared to the finite lattice model in Fig. \ref{tu4}(a), the total number of cells is kept constant by reducing and increasing the number of trivial phase unit cells by 5$\times$5 and 9$\times$9 respectively as in Fig. \ref{tu7}(a, b). We subjected the two finite lattices to spectral analysis and could observe an inward shift (Fig. \ref{tu7}(c)) and an outward shift (Fig. \ref{tu7}(d)) in the topological corner state. In practice, the topological corner state position can be adjusted by simply connecting the negative capacitance circuit to the corresponding piezoelectric sheet.

\section{conclusion}
In this paper, a honeycomb-shaped piezoelectric elastic plate is designed to achieve tunable topological corner states of the elastic wave HOTI. By connecting negative capacitance circuits to each of the inner or outer piezoelectric sheets in the unit cell, the topological phase is transformed accordingly. The band gap in the boundary state is present in supercell containing both trivial and non-trivial phase structures. The spectral analysis of its finite lattice reveals the presence of topologically protected corner states. Echoing the theory, the topological corner states in the hexagonal lattice were found to exist only at the 2$\pi$/3 corner. We designed both disturbed and vacant defect models and plotted the frequency energy spectrum curves. The strong robustness of the topological corner states is verified by comparison with the defect-free model. In addition, to highlight the tunability of the topological corner states of the piezoelectric elastic plate, we adjusted the piezoelectric parameter $\alpha$ to broaden the operating frequency range of the topological corner states. Switching the position of the negative capacitance circuit connection enables inward and outward shifts of the topological corner states, meaning that the topological corner states can appear at any position in the plane. The tunable topological corner state of this piezoelectric elastic plate enriches the study of topological physical phenomena in mechanical metamaterials. This study is of reference value in applications such as elastic wave energy localization, information transfer, and energy harvesting.
\section*{Acknowledgements}
This research is sponsored by the National Natural Science
Foundation of China (Grant nos. 12021002, 12072225, and 11991031). We thank Professor Yibin Fu at Keele University for helpful discussions and valuable advice.

\appendix
\section{Characterization of topological corner states}
By understanding the theory and methods in these articles\cite{chen21t,noh18t,ben19q}, we verify the topological corner properties of the piezoelectric elastic plate. Firstly the bulk polarisation here is expressed as
\begin{equation}
	\begin{aligned}
		\chi^{(6)}=(\boldsymbol{[M]},\boldsymbol{[K]})
	\end{aligned}
\end{equation}
where $\boldsymbol{[M]}$ and $\boldsymbol{[K]}$ are C$_{2}$ and C$_{3}$ invariants respectively. They are integers that can be expressed as
\begin{equation}
	\begin{aligned}
		\boldsymbol{[M]} =\# M_{1}-\# \Gamma_{1}^{(2)}
		\\ \boldsymbol{[K]} =\# K_{1}-\# \Gamma_{1}^{(3)}
	\end{aligned}
\end{equation}
where M$_{1}$ ($\Gamma_{1}^{(2)}$) is the number of energy bands with C$_{2}$ ($\pi$) rotation eigenvalue +1 below the band gap at point M ($\Gamma$) in the Brillouin zone. K$_{1}$ ($\Gamma_{1}^{(3)}$) is the number of energy bands with C$_{3}$ ($\pi/3$) rotation eigenvalue +1 below the band gap at point K ($\Gamma$) in the Brillouin zone. From Fig.\ref{tu8}, we observe that the eigenmodes of the three bands below the band gap are $\boldsymbol{[M]}$ = 0 for a trivial structure M$_{1}$ = 1 and $\Gamma_{1}^{(2)}$ = 1. For non-trivial structure M$_{1}$ = 1 and $\Gamma_{1}^{(2)}$ = 3, so $\boldsymbol{[M]}$ = 2 (ignoring the negative sign). The C$_{3}$ rotation operator and the chiral operator are swapped in this piezoelectric elastic plate model. It can be obtained that for both structures the invariant $\boldsymbol{[K]}$ = 0. We summarise the topological invariants in the two structures as follows
\begin{equation}
	(\boldsymbol{[M]},\boldsymbol{[K]})=
	\left\{
	\begin{array}{rcl}
		(0,0)       &      & {external\quad powering}\\
		(2,0)     &      & {internal\quad powering}.\\
	\end{array}
	\right.
\end{equation}
From the corner charge $Q_{corner}^{(6)}$=$\boldsymbol{[M]}/4$+$\boldsymbol{[K]}/6$ in the literature, it follows that
\begin{equation}
	Q_{corner}^{(6)}=
	\left\{
	\begin{array}{rcl}
		0       &      & {external\quad powering}\\
		1/2     &      & {internal\quad powering}.\\
	\end{array}
	\right.
\end{equation}
\begin{figure*}[!ht]
	\centering
	\renewcommand{\figurename}{Fig}
	\begin{minipage}{0.6\linewidth}
		\centering
		\includegraphics[scale=0.8]{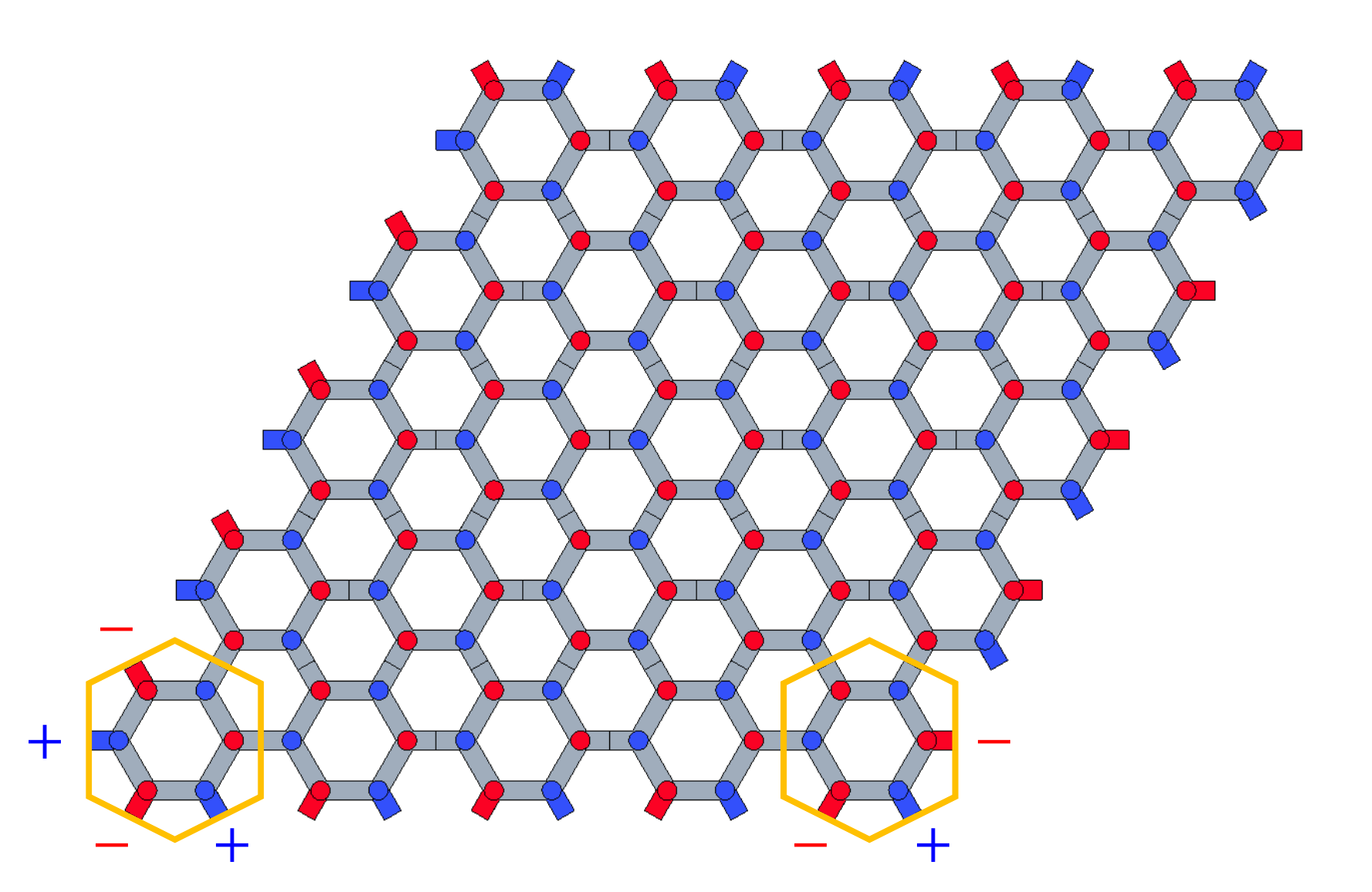}	
	\end{minipage}
	\hfill 
	\begin{minipage}[c]{.3\textwidth} 
		\centering
		\vspace{6cm}
		\caption{Topological corner modes at $\pi$/3 and 2$\pi$/3 in the rhombic finite lattice. The blue and red colors represent chiral charges with values of $+$1 and $-$1, respectively.}
		\label{tu9}
	\end{minipage}
\end{figure*}

To account for the difference in position between the emergence of topological and trivial corner states, the topological index $\mathcal N$ has been introduced to the characterization\cite{noh18t,fan19e}. The topological index $\mathcal N$ captures the interaction between the topology of the bulk Hamiltonian and the topology of the defect, representing the stable mode at the corners of the boundary. From Fig.\ref{tu9}, it can be seen that $N_{+}$=1 and $N_{-}$=1 at the edges not containing the corners of the rhombic finite lattice. At the $\pi$/3 corners, $N_{+}$= 2 and $N_{-}$= 2. Therefore the topological index at the edges and $\pi$/3 corners of the piezoelectric elastic plate is 0. While at the 2$\pi$/3 corners $N_{+}$=1, $N_{-}$=2, the topological index $\mathcal N$=1 means the stable mode occurs. Thus, the corner states at the 2$\pi$/3 corners of the boundaries of the different topological phase structures in the honeycomb elastic plate are topologically protected.
 
\nocite{*}

\bibliography{pz}

\end{document}